# Evolution of Thunderstorm Charge Structure Revealed by Particle Composition and Near-Surface Electric Field

**A.Chilingarian, B.Sarsyan**

**A A Alikhanyan National Lab (Yerevan Physics Institute)**

**Alikhanyan Brothers 2, Yerevan, Armenia. AM0036**

**Abstract**

Thunderstorm Ground Enhancements (TGEs) provide a direct particle-physics diagnostic of accelerating electric fields within thunderclouds. During 12–15 May 2026, Aragats detectors recorded a compact sequence of TGEs under closely related meteorological and electric-field conditions. Ten statistically resolved TGEs were selected for quantitative analysis using STAND3 and SEVAN Light gamma and electron proxy channels. The events were classified into negative (N), transition/disturbed negative (TD-N), and positive (P) states based on the stability of the near-surface electric field (NSEF). The results show that TGEs occur during both comparatively stable and rapidly evolving charge states. Stable negative events correspond to sustained exposure of the Main Negative (MN) charge region, whereas positive and TD-N events reflect Lower Positive Charge Region (LPCR) screening and rapid restructuring of the lower thundercloud dipole during End-Of-Storm Oscillation (EOSO)-like evolution. The analysis demonstrates that the electron/gamma composition depends primarily on the persistence and geometry of the lower-accelerating-field structure rather than solely on the instantaneous sign or magnitude of the NSEF. Rapid electric-field impulses, frequently associated with nearby lightning activity, mark active restructuring of the lower dipole and strongly influence electron transport to the detector level. All analyzed events occurred during persistent low-cloud-base conditions favorable for electron TGEs at Aragats. The May 2026 sequence provides a compact observational example linking EOSO behavior, LPCR evolution, lightning-related restructuring, and particle composition within a unified physical framework.

**Plain Language Summary**

Thunderstorms contain evolving layers of electric charge that can accelerate particles to high energies. These particles sometimes reach the ground, producing short-lived increases in radiation called Thunderstorm Ground Enhancements (TGEs). During 12–15 May 2026, detectors on Mount Aragats in Armenia recorded a sequence of ten TGEs under similar weather conditions. By combining particle measurements, electric-field observations, lightning data, and meteorological information, we show that TGEs reflect different stages of thunderstorm charge evolution. Some events occurred under stable negative electric-field conditions, while others developed during rapidly changing or predominantly positive field states. The particle mixture of electrons and gamma rays depends mainly on how stable and how close the accelerating electric field remains above the detector. The results support a physical picture in which the lower positive charge region inside the cloud evolves continuously during storm decay and lightning activity. TGEs therefore provide a way to study thunderstorm charge dynamics using particle detectors alongside traditional meteorological instruments.

**Key Points**

• Ten TGEs observed during May 2026 reveal stable negative, transitional, and positive charstages during the thunderstorm sequence.

• TGE composition depends on the persistence and geometry of the lower field structure rather than simply on NSEF polarity or TGE amplitude.

• Rapid NSEF impulses associated with lightning activity diagnose restructuring of the lower dipole during EOSO–LPCR evolution.

## 1. Introduction

This time-dependent charge evolution governs both lightning activity and the operation of high-energy atmospheric electron accelerators. Thunderstorm ground enhancements (TGEs, Chilingarian et al., 2010; 2011), detected at Aragats as increases in electron, positron, gamma-ray, and neutron fluxes, provide a direct observational probe of these electric-field structures. Unlike conventional meteorological measurements, TGE particle spectra carry information about the position, strength, and stability of the accelerating electric field in the lower part of the thundercloud. TGEs provide a direct particle-physics diagnostic of thundercloud charge dynamics. In a strong atmospheric electric field, seed electrons from the ambient cosmic-ray population can enter the runaway regime and initiate relativistic runaway avalanches (RREAs, Gurevich et al., 1992; Babich et al., 2001; Alexeenko et al., 2002; Dwyer et al., 2003), producing avalanches of electrons and gamma rays that are registered on the ground as TGEs. Because electrons attenuate rapidly after leaving the strong-field region, whereas gamma rays survive over larger distances, the measured electron/gamma composition is a sensitive indicator of the lower boundary and stability of the accelerating field above the detector.

The physical basis for this interpretation is the tripole structure of thundercloud charge. The classical Wilsonian dipole (Wilson, 1925) was substantially revised by Kuettner's Zugspitze measurements (Kuettner, 1950), which revealed a lower-charge region near the cloud base associated with graupel. This LPCR, together with the main negative charge region, forms the lower "graupel dipole," which is especially important for downward electron acceleration. Later studies of thunderstorm charge structure confirmed the broad relevance of the tripole concept and the importance of graupel charging, whose polarity depends strongly on temperature and riming conditions (Takahashi, 1978; Williams, 1989). Stolzenburg and Marshall (2008) emphasized that thunderstorm charge regions are highly dynamic structures whose evolution is strongly coupled to hydrometeor transport, lightning activity, and storm maturity. Their balloon soundings demonstrated that LPCR formation and decay are recurrent features of mature thunderstorms, especially during later stages of storm evolution. They also showed that the vertical ordering and temporal stability of charge layers can change rapidly on timescales comparable to TGE duration.

The importance of LPCR for TGE initiation was first explicitly demonstrated at Aragats by Chilingarian and Mkrtchyan (2012), who showed that the shape of NSEF disturbances can classify TGE initiation scenarios and that LPCR is not a secondary detail but an active participant in producing the strong lower electric field required for RREA. This result shifted the interpretation of TGEs from a simple "strong field above detector" picture to a dynamical model in which LPCR formation, screening, decay, and displacement control the observed particle flux.

Other lightning studies independently support the importance of the LPCR. Nag and Rakov (2009) showed that LPCR can facilitate different lightning types, and Tibetan Plateau observations also indicate that a large LPCR can modify the occurrence of negative cloud-to-ground flashes (Nag and Rakov, 2009; Qie et al., 2005). These studies link the LPCR not only to particle acceleration but also to lightning initiation and charge-neutralization pathways.

In this framework, two downward electron-accelerating structures emerge: the MN–MIRR dipole, formed between the main negative charge region and its mirror charge in the ground, and the MN–LPCR dipole. When both operate simultaneously, their electric fields can add coherently, exceeding the RREA threshold by tens of percent and producing intense TGEs. The same framework explains why a low cloud base and a low field-termination height are decisive for electron TGEs: gamma rays can traverse hundreds of meters of dense air, whereas electrons require a low-field boundary located tens of meters from the detector.

Weather Research and Forecasting (WRF, 2020) - based studies of TGE-producing clouds strengthened this interpretation by linking the lower, positively charged layer to graupel and the upper, negatively charged layer to snow (Chilingarian et al., 2021; Svechnikova et al., 2021). In analyzing Aragats events, the lower graupel cluster and the upper snow cluster formed a dipole-like charge structure with vertical separations of about 1–2 km, sufficient for avalanche multiplication. The lower positive layer had horizontal scales of several kilometers and charge densities of order 0.01–0.02 nC/m$^3$, making it physically capable of influencing both NSEF and the strong-field region where RREA develops.
The LPCR also influences lightning behavior. Larger or more mature LPCR structures can suppress negative cloud-to-ground flashes during part of storm development, favor inverted-polarity intracloud discharges, and create conditions in which TGEs are terminated by inverted IC or hybrid flashes. Thus, lightning distance and frequency indicate whether the local accelerating-field structure is stable or repeatedly disrupted by charge neutralization.
The winter-thunderstorm gamma-ray studies by Wada et al. (2021) provide another important perspective. Using atmospheric gamma-ray glows in Japanese winter thunderstorms, they demonstrated that long-duration radiation events are strongly linked to the evolution of thundercloud structure and lightning termination processes. Their work showed that gamma-ray glows frequently terminate abruptly after nearby lightning discharges, indicating direct coupling between relativistic particle acceleration and charge neutralization within the cloud. This interpretation closely parallels Aragats's observations, in which stable, electron-rich TGEs are associated with relatively weak lightning disturbance, whereas oscillatory positive/non-negative TGEs occur during periods of enhanced lightning activity and rapid restructuring of the accelerating-field structure. Wada et al. also emphasized the importance of low-altitude winter thunderclouds for efficient particle transport to ground level, reinforcing the broader conclusion that the distance between the detector and the lower boundary of the accelerating region is a critical parameter controlling the observed particle composition.
End-of-storm oscillations (EOSO) add the time dimension to this picture. In the EOSO scenario, the LPCR carried by graupel descends, screens the detector, weakens, and then exposes the detector to MN. The NSEF may shift from positive to deep negative and then recover toward fair-weather values as the tripole decays. The EOSO observations suggest that graupel fall, NSEF reversal, LPCR decay, and prolonged electron-rich TGEs can occur together. Recovered electron and gamma-ray spectra indicate that during such episodes, the lower boundary of the strong electric field can approach only tens of meters above the ground (Chilingarian et al., 2026).
Recent positron measurements add another diagnostic of LPCR evolution. During positive NSEF, the LPCR–ground dipole can decelerate electrons and accelerate positrons, producing enhanced 511 keV annihilation radiation. The 2024 Aragats positron studies showed that the emergence of LPCR can be traced through this annihilation channel, including a 500% enhancement of the 511 keV line during a strong positive-NSEF TGE. This confirms that particle composition can diagnose not only RREA intensity but also the polarity and vertical ordering of lower charge layers (Chilingarian et al., 2024a; 2024b).
The May 12–15, 2026 sequence extends this framework to a compact storm episode occurring under closely related meteorological conditions. Stable negative TGEs correspond to prolonged MN exposure and efficient downward electron transport. Positive TGEs correspond to LPCR screening, active charge rearrangement, or a late EOSO transition. The key diagnostics are therefore the sequence, duration, and stability of NSEF states, together with field-impulse activity, meteorological conditions, and electron/gamma composition. The experimental facilities of the Aragats research station and the TGE selection methodology were described in detail in Chilingarian et al. (2023; 2024c; 2024d; 2026).

## 2. Observed May 12-15 TGE sequence

May historically exhibits the highest TGE occurrence rates at Aragats. Mid-May 2026 was especially stormy, producing 10 significant TGEs within 3 days, reflecting the rapid evolution of the atmospheric electric field (AEF) above the station. The quantitative analysis of AEF is based on ten TGEs that are simultaneously significant in the STAND3 low-energy mixed electron-gamma channel ("1000" coincidence) and in the SEVAN Light gamma and electron channels ("11" and "01" coincidences). The events are numbered consecutively as TGE1-TGE10 in chronological order. This numbering is used consistently across all tables, figures, and discussions. Each event is characterized by its enhancement, NSEF evolution, cloud-base estimate, humidity, temperature, wind, rain, UV dose, and lightning distances.

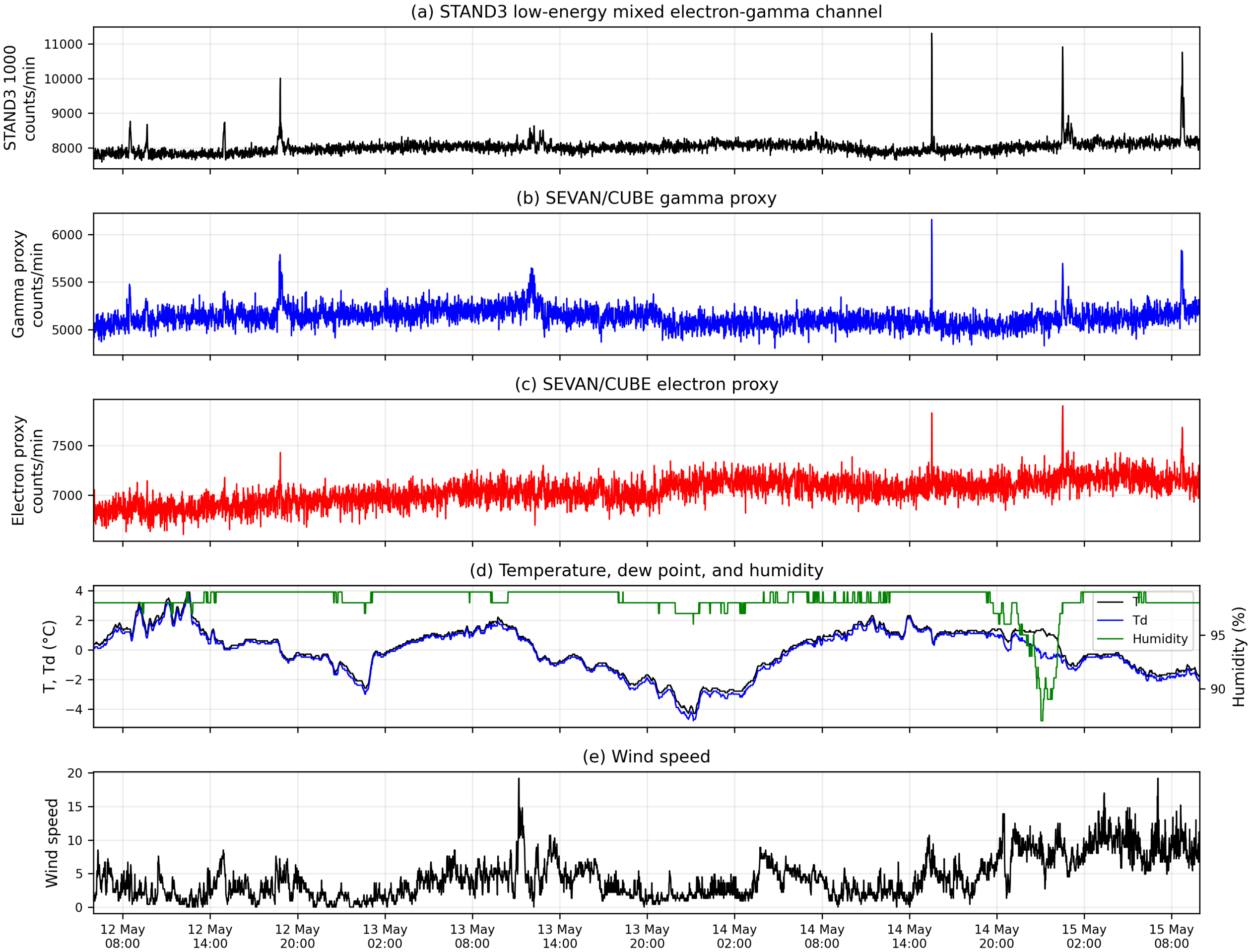

**Figure 1. Multichannel overview of the ten selected TGEs during 12-15 May 2026. The panels show STAND3 "1000" coincidence counts, SEVAN Light gamma and electron fluxes, temperature/dew point/humidity, and wind speed.**

Figure 1 is the observational backbone of the May 12-15 sequence. The particle enhancements are embedded within a continuous storm evolution. The first three panels show that the low-energy mixed channel, gamma proxy, and electron proxy respond differently from event to event. This separation means that the total particle enhancement and the electron/gamma composition convey different physical information. The STAND3 "1000" channel captures the overall low-energy enhancement, while the gamma and electron proxy channels help distinguish soft gamma-dominated responses from events with stronger charged-particle transport.

The meteorological panels show that the sequence developed in a saturated mountain-storm environment. Temperature and dew point remain close during most events, and humidity remains near saturation. This supports the interpretation that the cloud system remained near the detector altitude during the active period. Wind speed increases during the later stage on May 15, indicating a more dynamically developed storm environment.

TGE amplitudes vary widely in their gamma and electron responses, and the strongest low-energy enhancements are not necessarily associated with electron-rich events. Therefore, amplitude and electron/gamma composition are complementary diagnostics of changes in the thundercloud's charge structure.

In Figure 2, we present the relative enhancement of the 4 largest TGEs. For each event, the baseline is taken from a calm interval before TGE onset.

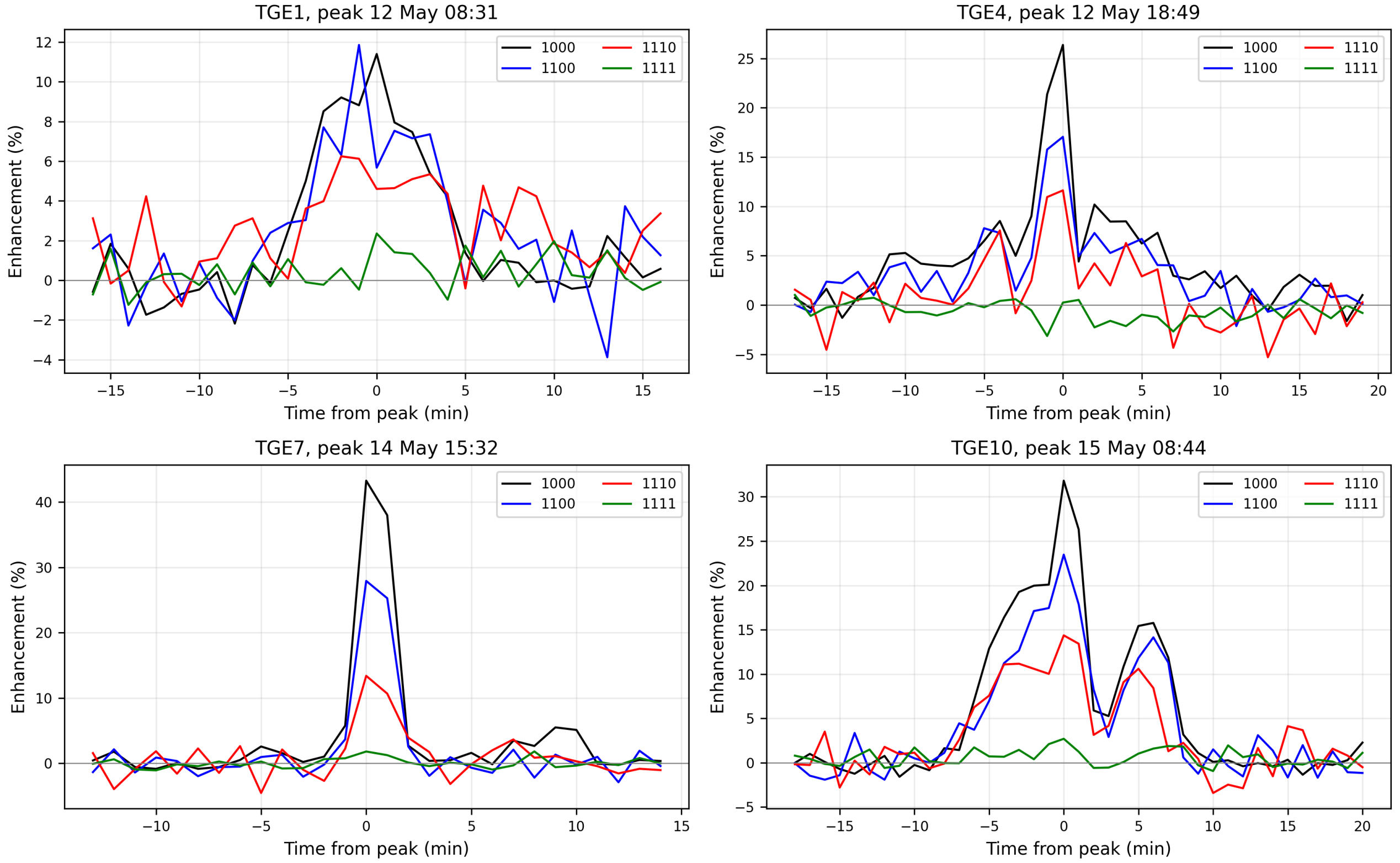

**Figure 2. The four largest TGEs extracted from the May 12-15 sequence. Each panel shows the STAND3 "1000", "1100", "1110", and "1111" coincidence enhancements relative to a calm pre-TGE baseline. The four channels correspond to progressively stricter penetration conditions through the STAND3 stack and therefore provide information on the hardness of the TGE particle flux.**

Figure 2 adds the energy-sensitive view that is not visible from the STAND3 1000 channel alone. The "1000" (>10 MeV) coincidence channel is the most inclusive and responds to the general low-energy electron-gamma enhancement. The higher coincidences, "1100" (>20 MeV), "1110" (>30 MeV), and "1111" (>40 MeV), require particles that penetrate deeper into the scintillator stack and are therefore more sensitive to the high-energy electron component. The relative behavior of the four curves provides a direct experimental indication of whether a TGE is soft or contains a stronger penetrating electron component.

The selected events show coherent enhancement across coincidence channels (except the highest "1111" channel), consistent with a comparatively stable lower accelerating structure that allows high-energy electrons to reach the detector. Thus, Figure 2 links detector response to thundercloud physics. The relative strengths of the four STAND3 coincidences indicate the depth and intensity of the particle flux, and this information must be interpreted together with the electron/gamma composition and the NSEF classification.

To compare the physical strength, statistical significance, and particle composition of the selected TGEs, Figure 3 presents the peak enhancement amplitudes together with gamma and electron proxy excesses derived from the STAND3 detector channels.

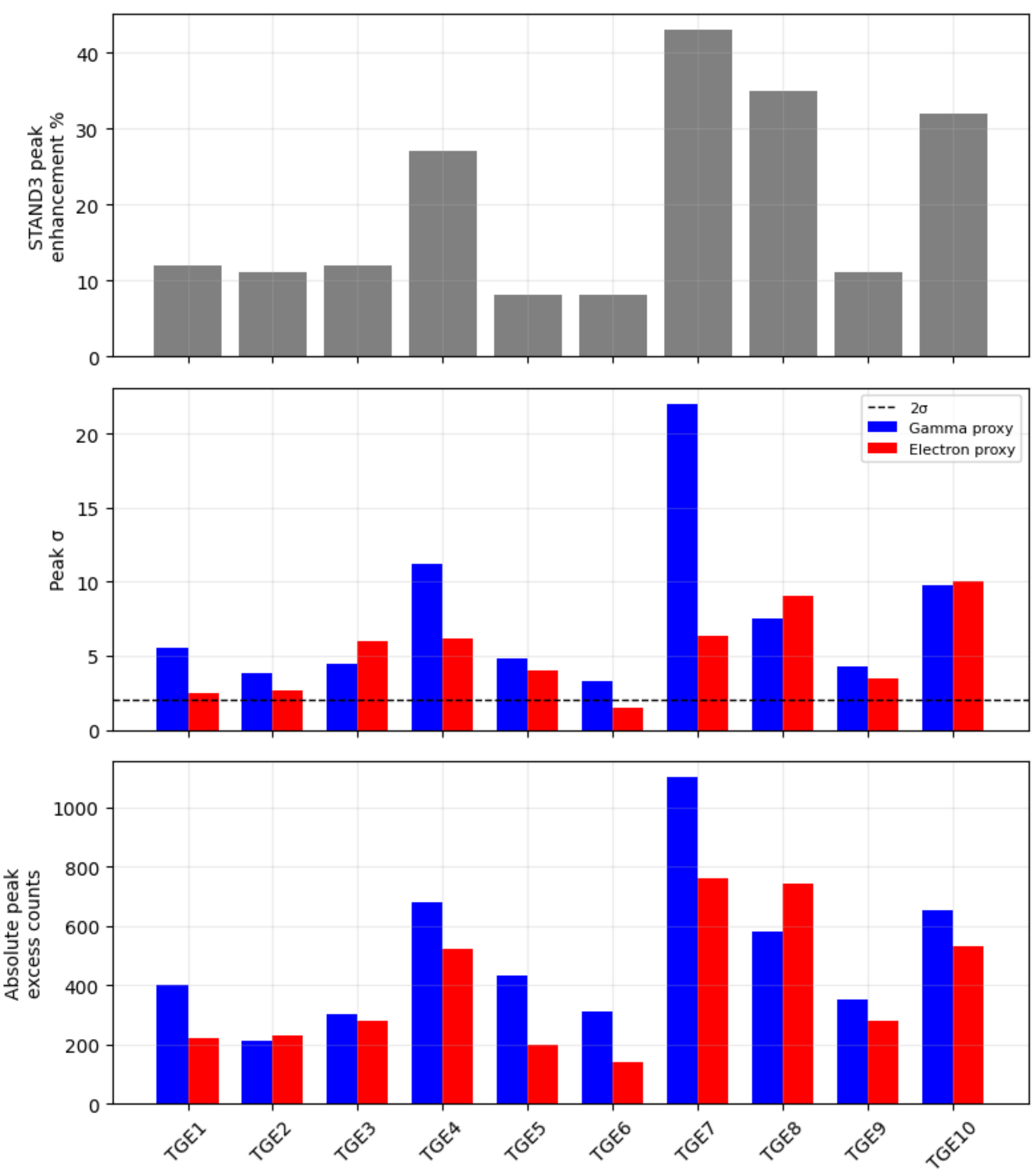

**Figure 3. Peak enhancement, statistical significance, and absolute excess for the ten selected TGEs. The upper panel shows the peak STAND3 enhancement, expressed as a percentage above baseline. The middle panel shows the statistical significance (σ) of the gamma and electron enhancements. The lower panel shows the corresponding absolute excess counts above the calm-weather baseline.**

STAND3 peak percent enhancement indicates the overall strength of the TGE. Statistical significance (σ) values test whether the gamma and electron enhancements are statistically resolved. Absolute excess gives the physical size of the detector response above the calm baseline. This separation prevents a weak but noisy event from being interpreted as electron-rich. The largest STAND3 enhancements occur in TGE7, TGE8, and TGE10. These high-amplitude events also show significant gamma and electron proxy responses, but their electron fractions remain moderate. In contrast, TGE2 and TGE9 have smaller or moderate STAND3 amplitudes but relatively large electron fractions.

## 3. Classification of TGE events by AEF structure

The polarity classification is based on the evolution of the near-surface electric field (NSEF) and the distribution of nearby lightning flashes. The goal of the classification is to identify physically distinct thundercloud configurations associated with TGE generation and termination. Figure 4 presents event-centered NSEF dynamics for all ten selected TGEs, together with nearby lightning flashes within 15 km of the detector site.

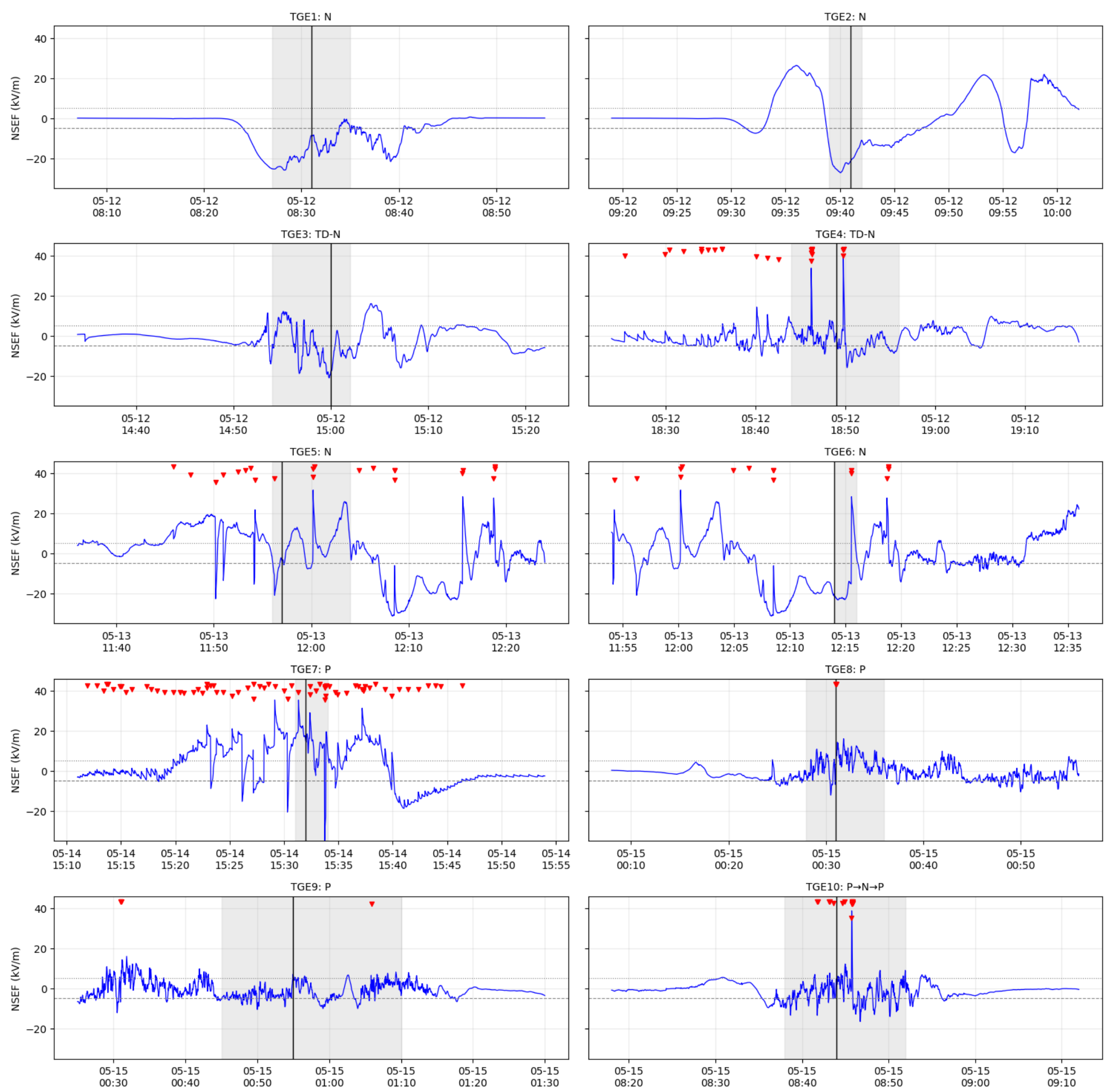

**Figure 4. Event-centered NSEF dynamics with nearby lightning flashes (<15 km). Blue curves show the NSEF evolution. Red triangles above the curves indicate nearby lightning flashes; higher**

**symbols correspond to smaller lightning distances. Vertical black lines mark the TGE peak. Dashed horizontal lines correspond to the −5 kV/m threshold used for polarity classification.**

In Figure 4, we see that most events contain negative NSEF intervals before the TGE peak. This behavior supports the interpretation that the negatively charged thundercloud region (MN) is directly exposed above the detector site during most of the analyzed TGEs. Abrupt upward jumps of the NSEF frequently occur immediately after the TGE peak and coincide with nearby lightning flashes that rapidly neutralize or restructure the lower dipole after a relatively stable negative field configuration had already produced the TGE. This effect is especially clear for TGE2 and TGE6, which are therefore classified as negative events. The disturbed negative (TD-N) events differ qualitatively from the negative events. In the TD-N group, the negative interval is interrupted by strong oscillations, impulsive field changes, or clusters of nearby lightning flashes occurring during the active TGE phase itself. Such behavior indicates ongoing restructuring of the lower thundercloud charge configuration during particle acceleration.

The group comprising TGE7, TGE8, and TGE9 belongs to the positive class (P) and displays distinct morphology. In these cases, the field remains oscillatory and predominantly positive/mixed while nearby lightning activity continues. This behavior is consistent with LPCR screening. TGE10 exhibits a particularly complex morphology: the first part of the event develops under predominantly positive NSEF conditions. After a nearby lightning flash, the field abruptly changes to negative values and then returns to positive values. Thus, TGE10 likely reflects rapid transitions between multiple charge configurations during thundercloud restructuring.

The classification derived from Figure 4, therefore, reflects physically distinct thunderstorm states that are often mixed. A negative TGE occurs during a stable negative NSEF state with weak impulse activity. A transition/disturbed negative TGE contains a negative interval but also shows field impulses or opposite-polarity excursions. A positive TGE is dominated by positive or mixed NSEF behavior and represents LPCR screening and active charge rearrangement.
Figure 5 shows the estimated distance from the detector level to the cloud base. However, the strong AEF region may not coincide exactly with the visible or thermodynamically estimated cloud base. The cloud-base distance was calculated from the surface temperature and dew-point spread using the approximate relation:

$$\text{cloud-base distance} \approx 122 \times (T - T_d)$$

where $T$ is the near-surface air temperature in °C and $T_d$ is the dew point in °C. This is a standard approximate estimate of the lifting condensation level.

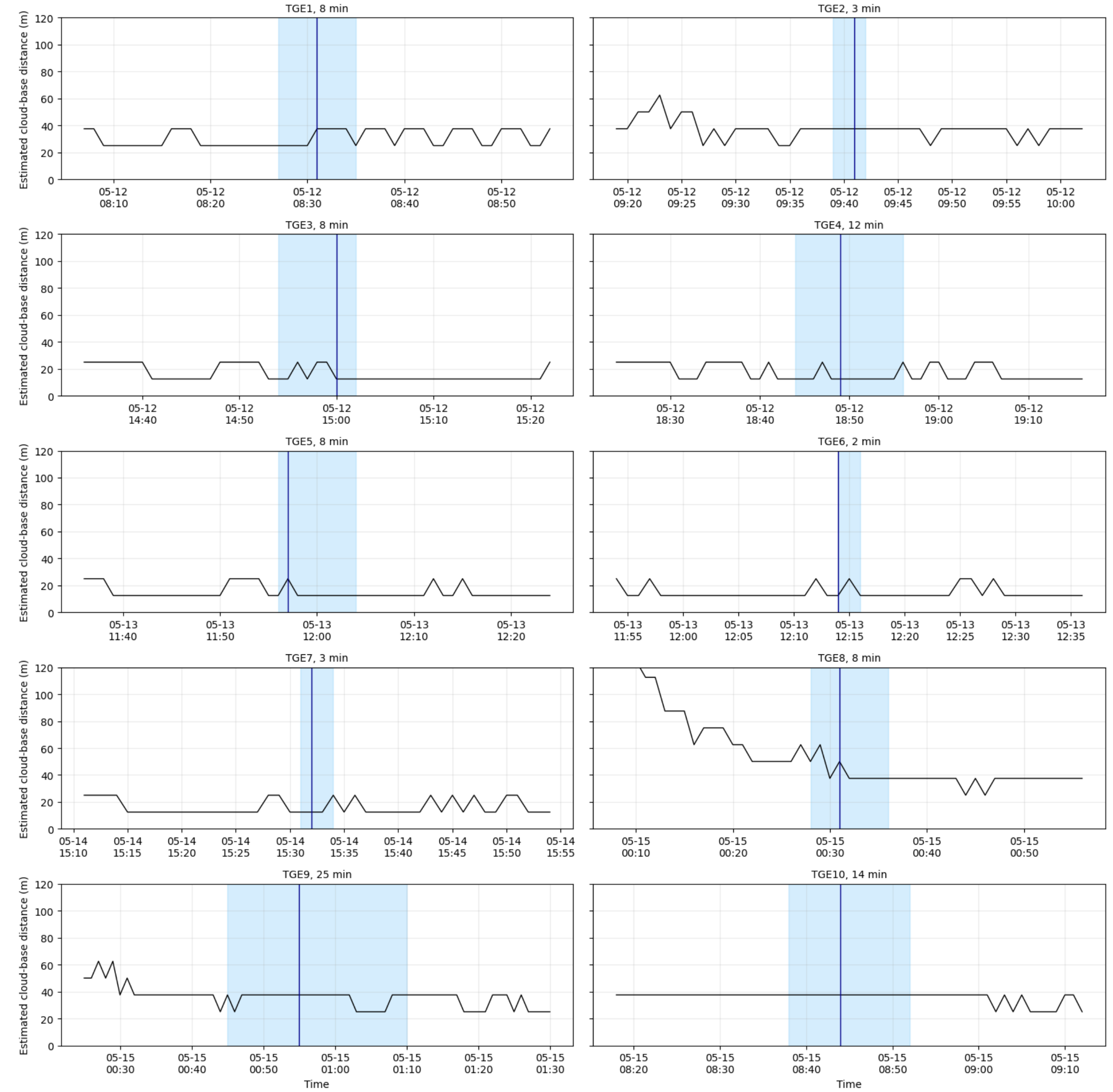

**Figure 5. Event-centered evolution of the estimated distance from the detector level to the cloud base for the ten analyzed TGEs. Each panel corresponds to one TGE. The blue curve shows the estimated cloud-base distance calculated from the temperature–dew-point spread.**

Figure 5 shows that across all events, the estimated cloud base was generally low, below 50 m, which is typical for May. Such low cloud-base conditions support both large TGE amplitudes and the detection of electron flux at ground level, because the distance between the lower cloud region and the detectors is small and atmospheric attenuation is reduced. Thus, Figure 5 provides supporting meteorological evidence for favorable electron TGE conditions.

In Table 1, we quantitatively summarize the diversity of electric-field configurations observed during the May 12–15 thunderstorm episode. The coexistence of long negative intervals, short positive intervals, and impulsive field changes indicates rapid restructuring of the lower thundercloud charge system during TGE production.

**Table 1. Event-level classification for the ten selected TGEs. Negative and positive intervals denote the longest continuous periods during which the NSEF remained below −5 kV/m and above +5 kV/m, respectively. NSEF impulses are 1-s field jumps larger than 10 kV/m.**

| Event | Start | Peak | Dur. min | Class | Neg. sec | Pos. sec | NSEF impulses | \|NSEF\|max | Cloud base m | electr. fraction |
|---|---|---|---|---|---|---|---|---|---|---|
| TGE1 | 2026-05-12 08:27 | 2026-05-12 08:31 | 8 | N | 413 | 0 | 0 | 25.9 | 29.8 | 0.224 |
| TGE2 | 2026-05-12 09:39 | 2026-05-12 09:41 | 3 | N | 181 | 0 | 0 | 27.2 | 36.6 | 0.448 |
| TGE3 | 2026-05-12 14:54 | 2026-05-12 15:00 | 8 | TD-N | 114 | 86 | 0 | 21.1 | 16.3 | 0.414 |
| TGE4 | 2026-05-12 18:44 | 2026-05-12 18:49 | 12 | TD-N | 135 | 11 | 2 | 40.7 | 14.1 | 0.236 |
| TGE5 | 2026-05-13 11:56 | 2026-05-13 11:57 | 8 | N | 44 | 123 | 4 | 31.5 | 13.6 | 0.099 |
| TGE6 | 2026-05-13 12:14 | 2026-05-13 12:14 | 2 | N | 93 | 22 | 2 | 28.2 | 16.3 | 0.319 |
| TGE7 | 2026-05-14 15:31 | 2026-05-14 15:32 | 3 | P | 7 | 107 | 8 | 35.3 | 15.2 | 0.397 |
| TGE8 | 2026-05-15 00:28 | 2026-05-15 00:31 | 8 | P | 17 | 57 | 0 | 16.0 | 42.0 | 0.501 |
| TGE9 | 2026-05-15 00:45 | 2026-05-15 00:55 | 25 | P | 77 | 23 | 0 | 10.5 | 33.8 | 0.445 |
| TGE10 | 2026-05-15 08:38 | 2026-05-15 08:44 | 14 | N-P-N* | 33 | 20 | 2 | 38.7 | 36.6 | 0.405 |

* Because the TGE10 developed primarily during the initial negative interval, it is included in the negative class despite its later polarity reversals.

The electron fraction shows no dependence on NSEF polarity. Significant electron fractions are observed for both negative and positive events. This result supports the interpretation that the near-surface electric field alone does not uniquely determine the particle composition of the TGE flux; a low cloud base is a necessary condition for electron detection, while charge-structure evolution and field stability control event-to-event differences. The estimated cloud-base distances remain generally low throughout the analyzed sequence, generally below about 50 m, with most events between 15 and 40 m**.** These values support favorable conditions for reduced atmospheric attenuation and efficient transport of high-energy electrons and gamma rays to the detector level.

**4. Meteorological environment of the selected TGEs**

**Table 2. Meteorological and electric-field conditions for each selected TGE. Cloud base is estimated from the temperature-dew-point spread. NSEF impulses are 1-s field jumps larger than 10 kV/m.**

| Event | Peak | T C | Td C | RH % | Cloud base m | Wind | Rain | UV | \|NSEF\|max | NSEF mean | NSEF impulses |
|---|---|---|---|---|---|---|---|---|---|---|---|
| TGE1 | 2026-05-12 08:31:00 | 1.28 | 1.03 | 98.0 | 29.8 | 3.58 | 0.000 | 0.013 | 25.9 | -15.5 | 0 |
| TGE2 | 2026-05-12 09:41:00 | 1.12 | 0.825 | 98.0 | 36.6 | 1.10 | 0.000 | 0.033 | 27.2 | -21.4 | 0 |
| TGE3 | 2026-05-12 15:00:00 | 0.222 | 0.089 | 99.0 | 16.3 | 6.61 | 0.000 | 0.000 | 21.1 | -4.43 | 0 |
| TGE4 | 2026-05-12 18:49:00 | 0.023 | -0.092 | 99.0 | 14.1 | 4.62 | 0.000 | 0.000 | 40.7 | -2.59 | 2 |
| TGE5 | 2026-05-13 11:57:00 | 0.500 | 0.389 | 99.0 | 13.6 | 2.74 | 0.000 | 0.000 | 31.5 | 4.65 | 4 |
| TGE6 | 2026-05-13 12:14:00 | 0.067 | -0.067 | 99.0 | 16.3 | 1.63 | 0.000 | 0.000 | 28.2 | -13.0 | 2 |
| TGE7 | 2026-05-14 15:32:00 | 0.950 | 0.825 | 99.0 | 15.2 | 6.82 | 0.000 | 0.000 | 35.3 | 13.4 | 8 |
| TGE8 | 2026-05-15 00:31:00 | -0.444 | -0.789 | 97.7 | 42.0 | 8.59 | 0.000 | 0.000 | 16.0 | 3.82 | 0 |
| TGE9 | 2026-05-15 00:55:00 | -0.942 | -1.22 | 98.0 | 33.8 | 6.12 | 0.000 | 0.000 | 10.5 | -1.90 | 0 |
| TGE10 | 2026-05-15 08:44:00 | -1.43 | -1.73 | 98.0 | 36.6 | 10.7 | 0.800 | 0.007 | 38.7 | -2.00 | 2 |

Table 2 shows that several events include long negative intervals, accompanied by positive excursions or rapid impulses, indicating that particle acceleration occurred during active restructuring of the lower thundercloud charge system. The electron fraction varies substantially across events and does not uniquely correlate with either TGE amplitude or instantaneous NSEF polarity. For example, TGE8 and TGE9 exhibit large electron fractions under predominantly positive or mixed conditions, whereas several strongly negative events show smaller electron contributions. The impulse statistics indicate that rapid field restructuring occurs on timescales comparable to those of TGE evolution. Therefore, lightning-related field changes and transient charge rearrangements are part of the charge configuration that controls TGE production rather than secondary perturbations superposed on otherwise stationary events.

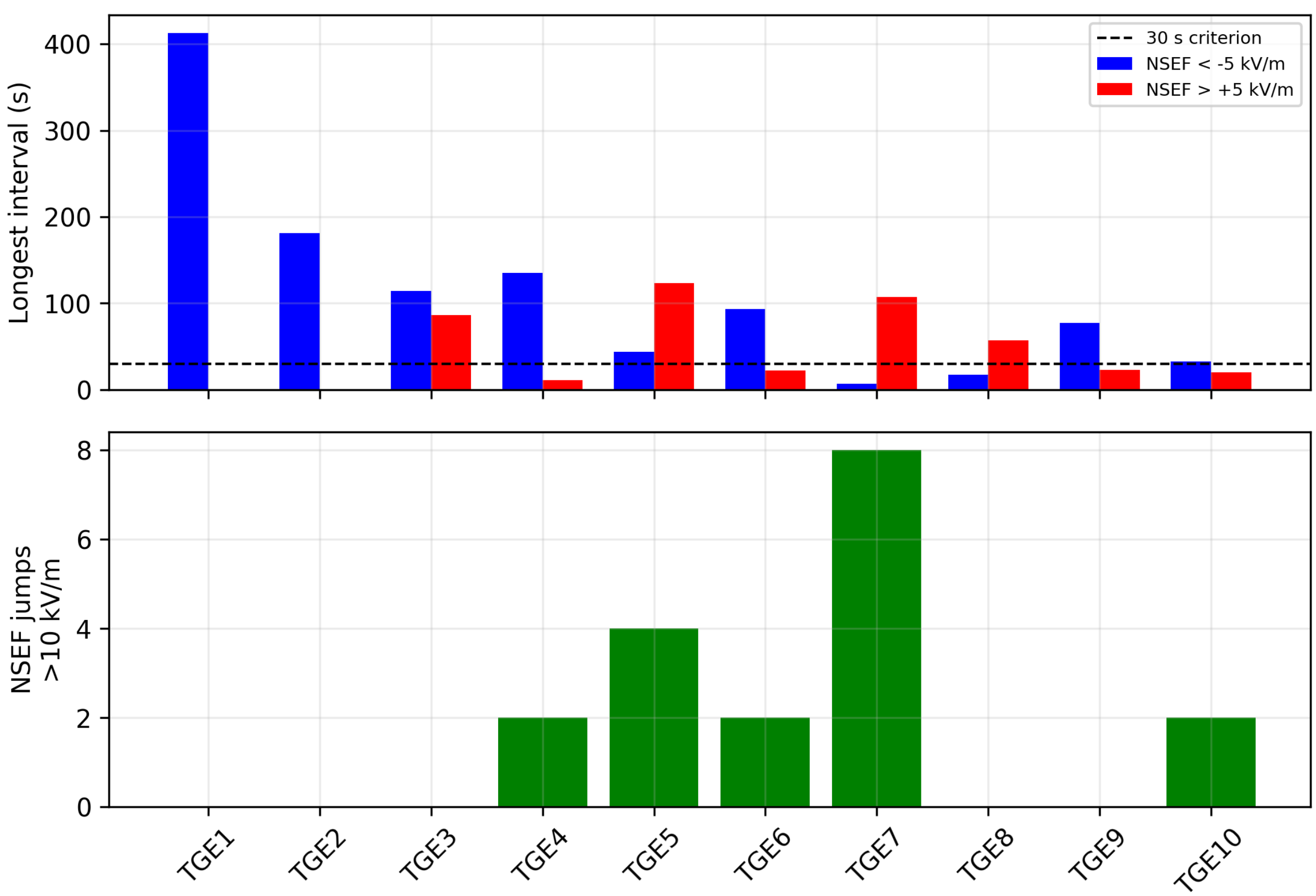

**Figure 6. Second-scale polarity stability and rapid electric-field impulses. The upper panel compares the longest continuous intervals with NSEF < −5 kV/m and NSEF > +5 kV/m. The dashed line marks the 30-s stability criterion. The lower panel shows the number of 1-s NSEF jumps larger than 10 kV/m.**

Figure 6 illustrates the adopted classification scheme. Negative (N, blue) events are characterized by relatively long continuous negative intervals of NSEF and weak impulse activity. These events correspond to comparatively stable exposure of the main negative (MN) thundercloud charge region above the detector. However, even these events are not entirely static and can still terminate due to rapid lightning-induced field rearrangement. Transition/disturbed negative (TD-N) events satisfy the negative-state criterion but also contain positive excursions or impulsive field changes. These events represent rapidly evolving thundercloud configurations in which the lower negative region remains partially exposed while the local charge structure simultaneously undergoes restructuring. Positive (P, red) events occupy the most dynamically variable part of the sequence. They are characterized either by dominant positive intervals or by strong field impulses accompanied by rapidly changing polarity conditions.
Previous classification introduced four phenomenological classes of TGEs based on large-scale patterns of near-surface electric-field evolution (Chilingarian and Mkrtchyan, 2012), including polarity reversals, abrupt decreases in the electric field, and multiple disturbances accompanied by lightning activity. The present study refines it on much shorter temporal scales. Instead of classifying the entire thunderstorm episode based on its large-scale electric-field evolution, we classify individual TGEs according to second-scale polarity stability, lower dipole instability, and transient restructuring of the near-surface electric field.

The lower panel shows that rapid field impulses track charge restructuring on timescales comparable to those of TGE development. Such impulses can accompany nearby lightning activity or the rearrangement of the lower cloud charge layers during EOSO. EOSO features repeated polarity reversals and oscillatory behavior of the NSEF during the decay stage of thunderstorms, associated with evolving thundercloud charge regions and screening layers (Marshall et al., 2009; Pawar and Kamra, 2007). Whereas classical EOSO studies focused on large-scale polarity oscillations during storm decay, the present analysis resolves impulsive transitions of charge configuration that occur simultaneously with particle acceleration.

**Table 3. Class-level averages for the ten selected TGEs using the revised dynamical classification.**

| Class | N | Mean amp. % | Mean dur. min | Mean \|NSEF\| | Mean hieght m | Mean impulses | Mean electron fraction |
|---|---|---|---|---|---|---|---|
| negative (N) | 4 | 14.4 | 5.25 | 24.0 | 24.1 | 1.50 | 0.273 |
| positive (P) | 3 | 29.3 | 12.0 | 22.8 | 30.3 | 2.67 | 0.448 |
| transition/disturbed negative (TD-N) | 3 | 26.6 | 11.3 | 33.5 | 22.3 | 1.33 | 0.352 |

Table 3 summarizes the average properties of the revised event classes. The TD-N class exhibits the largest mean electric-field magnitude, reflecting strongly disturbed charge configurations. Negative (N) events correspond to the most stable accelerating-field configurations. Positive (P) events exhibit the largest mean STAND3 flux enhancement and the highest mean electron fraction. However, because only one of the three positive events contained strong NSEF impulses, the average impulse rate for this class should not be regarded as a robust distinguishing characteristic.

The revised classification, therefore, emphasizes the dynamical evolution of thunderclouds rather than static electric-field polarity alone. This represents the principal difference between the present analysis and the earlier 2012 classification scheme.

## 5. Discussion and conclusions

The combined analysis of the May 12–15, 2026, TGE sequence, including NSEF dynamics, lightning activity, particle composition, and meteorological conditions, demonstrates that TGE development is controlled by the evolution of the lower thundercloud charge structure.

The principal result of this study is the emergence of a three-state dynamical classification comprising negative (N), transition/disturbed negative (TD-N), and positive (P) events. Unlike simplified polarity-based schemes, the revised classification is based on second-scale NSEF stability, rapid field impulses, and resolved electron/gamma composition. The analyzed sequence contains stable negative, transitional, and positive charge states in comparable proportions, indicating continuous evolution of the lower thundercloud charge structure during the May storm episodes.

Within the EOSO–LPCR framework, LPCR partially screens the lower thundercloud dipole, while the MN charge region remains the primary source of downward-directed acceleration. As the LPCR evolves, weakens, or shifts, the detector alternately samples LPCR-screened, mixed, and MN-dominated configurations. The TD-N events represent intermediate states in this evolution because they contain prolonged negative intervals alongside impulsive restructuring or short positive excursions. These events, therefore, record active evolution of the lower charge structure rather than stationary polarity states.

Negative events correspond to the most stable MN-exposure regime. Positive events occur when LPCR screening dominates the surface field or when the lower dipole is restructuring during late EOSO evolution.

The analysis further shows that the electron/gamma mixture depends primarily on the geometry of the accelerating structure and the height of the AEF lower boundary. Rapid NSEF impulses indicate active restructuring of the lower dipole, often associated with nearby lightning activity. Gamma rays produced in the avalanche can withstand such disturbances more easily than electrons, which lose energy outside the strong-field region.

The meteorological observations indicate that all analyzed events occurred within a persistent low-cloud-base regime marked by high humidity. These conditions are necessary for ground-level electron detection at Aragats. For positive events, the LPCR position determines the lower boundary and geometry of the MN–LPCR accelerating region. Consequently, the observed electron fraction is controlled mainly by the strength and vertical extent of the MN–LPCR electric field, together with the distance electrons must travel after leaving the strong-field region.

The combined analysis demonstrates that TGEs can serve as probes of thundercloud charge structure. Stable negative intervals indicate sustained MN exposure, positive intervals reflect LPCR screening or late-stage EOSO restructuring, and rapid impulses reveal active reorganization of the lower charge system. The resolved electron/gamma composition then provides information on the persistence and geometry of the accelerating structure.

Overall, the study shows that a multi-parameter approach combining particle detectors, NSEF measurements, lightning observations, and meteorological data provides a substantially informative description of thunderstorm evolution. The analyzed sequence presents a compact observational example linking EOSO behavior, LPCR evolution, lightning-related restructuring, and electron-rich TGEs within a unified physical framework.

**Acknowledgements**

The authors acknowledge the support of the Science Committee of the Republic of Armenia (Research Project No. 21AG-1C012).

**Data Availability Statement**

The full set of TGEs is available in both numerical and graphical formats at the cosmic ray division's database of the Yerevan Physics Institute: http://crd.yerphi.am/adei/. The WIKI section of the database contains instructions on selecting appropriate detectors and times, as well as on uploading data.

**Conflict of Interest**

The authors declare no conflicts of interest relevant to this study.

**Author Contributions**

A.C. conceived the study and led the analysis. B.S. contributed to data processing and detector operation. All authors discussed the results and contributed to the final manuscript.

**Glossary: definitions and terminology used in this analysis**

| Term | Meaning |
|---|---|
| RREA | Relativistic Runaway Electron Avalanche |
| TGE | Thunderstorm Ground Enhancement |
| NSEF | Near-surface electric field. Its sign is a ground-level projection of cloud charge structure. |
| LPCR | Lower positive charge region, commonly associated with positively charged graupel in the lower part of the thundercloud. |
| MN | Main negative charge region of the thundercloud. Exposure to MN is usually expressed by prolonged negative NSEF. |
| EOSO | End-of-storm oscillation: rapid polarity changes and recovery of NSEF during storm decay, associated with descent and weakening of charged layers. |
| Accelerating-field structure | The spatial and temporal arrangement of LPCR, MN, upper positive charge, and the strong electric field where RREA develops; it includes the lower boundary of the field and its stability against lightning-related restructuring. |
| Negative TGE | A TGE occurring during a stable negative NSEF state with weak impulse activity. |
| Transition/disturbed negative TGE | A TGE with a negative interval but with field impulses or polarity excursions indicating continuing charge rearrangement. |
| Positive TGE | A TGE dominated by positive or mixed NSEF behavior, associated with LPCR screening, active charge rearrangement, or late EOSO state. |

**Table 6. Definitions used throughout the analysis.**